\documentclass{nature}
\usepackage{epsfig,psfrag,graphics,color,verbatim}

\title{The moment of truth for WIMP Dark Matter}

\author{Gianfranco Bertone$^{1}$}

\def\crosssection{\sigma_{\chi-\mbox{\tiny{p}}}^{\mbox{\tiny{SI}}}}

\def\lesssim{\buildrel < \over {_{\sim}}}

\begin{document}

\maketitle

\begin{affiliations}
 \item Institut d'Astrophysique de Paris, UMR 7095-CNRS, Universit\'e Pierre et Marie Curie, 98 bis Boulevard Arago 75014, Paris, France \& Institute f\"ur Theoretische Physik, Universit\"at Z\"urich, Winterthurerstrasse 190, CH-8057 Z\"urich, Switzerland
\end{affiliations}

\begin{abstract}
We know that dark matter constitutes 85\% of all the matter in the Universe, but we do not know of what it is made. Amongst the many Dark Matter candidates proposed, WIMPs (weakly interacting massive particles) occupy a special place, as they arise naturally from well motivated extensions of the standard model of particle physics. With the advent of the Large Hadron Collider at CERN, and a new generation of astroparticle experiments, the moment of truth has come for WIMPs: either we will discover them in the next five to ten years, or we will witness the inevitable decline of WIMPs.
\end{abstract}

The foundations of the modern Dark Matter problem\cite{book} have been laid in the 70's and 80's, after decades of slow accumulation of evidence \cite{Einasto:2009zd}. Already in the 30's it was noticed that the Coma Cluster seemed to contain much more mass than what could be inferred from visible galaxies \cite{Zwicky:1933gu}, and a few years later, it became clear that the Andromeda galaxy rotates very fast at large radii, {\it as if} most of its mass lied in its outer regions \cite{babcock}. Several other pieces of evidence provided further support to the Dark Matter hypothesis, including the so called timing-argument \cite{kahn,einasto}, until in the 70's rotation curves were extended to larger radii and to many other spiral galaxies, proving the presence of large amounts of mass on scales much larger than the size of galactic disks \cite{bosma,rubin}. 
Although it is in principle possible to explain these observations in terms of new theories of gravity \cite{modified} (after all, we only have gravitational evidence for Dark Matter), lensing observations of galaxy clusters provide a formidable challenge to these theories \cite{BULLET1,BULLET2}.

Today, we have entered in the era of {\it precision Cosmology}: we can determine the abundance of Dark Matter in the Universe with exquisite accuracy \cite{Komatsu:2010fb}; we have a much better understanding of how Dark Matter is distributed in structures ranging from Dwarf Galaxies to Clusters of galaxies, thanks to high-resolution numerical simulations made possible by modern supercomputers \cite{DiemandMoore}, and to lensing observations \cite{Mellier}; we even have a rather precise idea of how the Milky Way has formed, and of what the {\it local} abundance of Dark Matter is \cite{catena09,pato10}. What is more important, we know today that Dark Matter cannot be made of ordinary matter, so new particles {\it must} exist \cite{Taoso:2007qk}, unless we are completely misled by a wide array of astrophysical and cosmological observations. 

Particle physicists have proposed literally tens of possible Dark Matter candidates. Axions, for instance, are hypothetical particles whose existence was postulated to solve the so called strong CP problem in Quantum Chromo Dynamics, and they are known to be very well motivated Dark Matter candidates \cite{AXIONREVIEW,axion2}. Other well-known candidates are Sterile Neutrinos, which interact only gravitationally with ordinary matter, apart from a small mixing with the familiar neutrinos of the Standard Model \cite{STERILE,sterile2}. A wide array of other possibilities have been discussed in the literature, and they are currently being searched for with a variety of experimental strategies \cite{Bergstrom,Bertone:2004pz}.

The most studied class of candidates, however, is that of WIMPs (for weakly interacting massive particles), which have the virtue of naturally achieving the correct relic abundance in the early Universe. The reason why they became so popular is that WIMP candidates arise naturally from theories that seek to extend the Standard Model of particle physics, and to embed it in a more fundamental theory. In particular, it was noticed back in 1983 that one of the most promising extensions of the Standard Model, Supersymmetry, provides an excellent Dark Matter candidate: the neutralino \cite{Goldberg:1983nd,Ellis:1983ew,blumenthal,Jungman:1995df}. This particle fulfills all the properties of the good Dark Matter candidate, and it has become over the years a prototypical example of WIMPs. Its mass can range from about 50 GeV to a few TeV, and its interaction cross section with ordinary matter and with itself are such that it can account for all the Dark Matter in the Universe while still remaining consistent with all known experiments.

\paragraph{BOX 1: WIMPs.} {\it In the most simple WIMP models, Dark Matter particles are kept in thermal and chemical equilibrium in the early Universe with all other particles, by virtue of their self-annihilation into particles of the Standard Model and vice versa. Their density rapidly decreases as the Universe expands, until it becomes so low that WIMPs cannot self-annihilate anymore and they freeze-out from equilibrium, i.e. their comoving number density remains fixed. Under some simplifying assumptions  \cite{Bertone:2004pz}, the relic abundance of WIMPs in the Universe $\Omega_\chi$ can be simply expressed in terms of the self-annihilation cross section $\sigma v$}
\begin{equation}
\Omega_\chi \approx \frac{3 \times 10^{-27} {\rm cm}^3 {\rm s}^{-1}}{\sigma v}
\end{equation}
{\it Since the measured value of $\Omega_\chi$ is around 0.1 \cite{Komatsu:2010fb}, the self-annihilation cross section which is required in order to achieve the appropriate relic density is $\sigma v \approx 3 \times 10^{-26} {\rm cm}^3 {\rm s}^{-1}$, a cross section typical of Weak interactions in the Standard Model, hence the name WIMPs. Although in this simplified calculation the relic density does not depend strongly on the mass of the Dark Matter particle, $m_\chi$, the maximum and minimum annihilation cross sections of the most common candidates does depend on it, therefore constraining the values of $m_\chi$ to the range $10 \lesssim (m_\chi / {\rm GeV})  \lesssim  10^5 $. \cite{Bertone:2004pz}}

If Dark Matter is made of WIMPs, we should be able to detect it. We could in principle observe the interaction of Dark Matter particles with nuclei in underground detectors, as proposed back in 1985 \cite{GoodWitt}, or we may detect the products of annihilation or decay of these particles, as first discussed almost 3 decades ago \cite{SilkSrednicki,Silk:1985ax,Bertone:2004pz}. Although all the search strategies so far devised have failed to provide incontrovertible evidence for Dark Matter particles, today a new generation of particle astrophysics experiments is about to start, or has already started taking data. Furthermore, the Large Hadron Collider at CERN has recently started operations, and it is expected to find, or to severely constrain, the most studied extensions of the Standard Model, including Supersymmetry.

I argue here that the moment of truth has therefore come for WIMP Dark Matter, for we will either discover them at the LHC and in particle astrophysics experiments in the next 5 to 10 years, or the the case for WIMPs will become weak, and we will witness the inevitable decline of the WIMPs.

\section{Indirect detection}
Indirect detection consists in the search for the annihilation or decay products of Dark Matter particles, such as photons, anti-matter and neutrinos. WIMPs in fact are expected to annihilate efficiently in regions where they accumulate, such as the center of galactic halos, or substructures such as dwarf galaxies, since the annihilation rate depends on the number density {\it squared}. Once they annihilate, they produce secondary particles, such as quarks and gauge bosons, which subsequently fragment and decay in the aforementioned final states. The typical energy of these final states is about a tenth of the Dark Matter particle mass, so we can search indirectly for Dark Matter by looking for an excess of photons, anti-matter or neutrinos in astrophysical data at energies between 1 GeV and 10 TeV (see Box 1).  

Although in principle interesting, obtaining convincing evidence from astrophysical observations has proven a very difficult task. It is in fact easy to fit almost any excess in the measured energy spectrum of photons or anti-matter, at any energy, in terms of Dark Matter particles with suitable properties. One simply has to follow three easy steps: {\it i)} adjust the normalization of the flux by changing the distribution of Dark Matter particles and their annihilation cross section; {\it ii)} choose a Dark Matter mass that provides the correct energy scale and {\it iii)} fit the spectral features by choosing an appropriate annihilation channel and, in the case of anti-matter, by tuning the propagation parameters. In practice, one has enough freedom to fit almost any astrophysical observation, and in fact features in the data of many experiments of the last 5-10 years, have been tentatively interpreted in terms of different Dark Matter candidates, sometimes even at the cost of making unrealistic assumptions on the nature and distribution of Dark Matter. 

The most recent example is the rise in the energy spectrum of the positron ratio measured by PAMELA above 10 GeV \cite{Adriani:2008zr}. The standard WIMP model (i.e. a particle with a mass in the $10^2-10^3$ GeV range and a {\it thermal} cross section, $\sigma v \sim 10^{-26}$ cm$^3$ s$^{-1}$) can hardly account for this feature, so new ad-hoc candidates have been proposed: particles with a very large annihilation cross section (high enough to match the normalization of the positron ratio, but not too much, in order to avoid cosmological constraints \cite{cmb1,cmb2}), annihilating only to leptons (to evade anti-protons constraints \cite{antip1}), and with a density profile shallower than what suggested by numerical simulations (to evade gamma-ray constraints from the Galactic center \cite{gamma1}). There is therefore a possible combination of parameters that can be made compatible with all observations, but this is certainly not enough to claim discovery of Dark Matter, for there are less exotic astrophysical sources that can account for the same feature without invoking new particles with ad-hoc properties. 

Fortunately, there are actually a number of astrophysical observations that might lead to {\it convincing} evidence, in the sense that they could be explained {\it only} in terms of Dark Matter, while being incompatible with a standard astrophysical interpretation. A typical example of smoking-gun evidence is the observation of a high-energy gamma-ray line, that would point directly to the existence of new particles annihilating directly to photons. In fact, if WIMPs do not produce photons through the fragmentation and decay of secondary particles, but directly, the photons produced in the annihilation will be mono-energetic, thus producing a line in the gamma-ray spectrum at an energy equal to the mass of the Dark Matter particle. The Fermi LAT satellite has however put stringent constraints on the possibility to observe of lines \cite{Abdo:2010nc}, excluding annihilation cross section to photons higher than the thermal cross section, and we can expect an improvement in sensitivity of one order of magnitude at most over the next decade (at least in the energy range where the sensitivity is limited by statistics, and not by the background, in which case the sensitivity scales with the square root of time). 

Another very clean signature of Dark Matter annihilations would be the observation of high-energy neutrinos from the center of the Sun \cite{Silk:1985ax}. Solar neutrinos produced in nuclear reactions have in fact energies in the MeV range, so the observation of $10^2$ - $10^4$ GeV neutrinos would require an explanation in terms of new physics, and the well studied process of capture and annihilation of Dark Matter particles in the Sun would provide it. The problem is that the neutrino telescope IceCube, currently under construction at the South Pole, so far has not found any evidence for an excess of neutrinos from the Sun, and in the next 5 years, the experiment will improve its sensitivity only by a factor of $\sim 5$, and extend the threshold down to 50 GeV, with the construction of a more densely instrumented portion of detector, called DeepCore \cite{Halzen:2009vu}. Even with this technical improvements and longer exposure, most of the Supersymmetry parameter space will remain inaccessible, and the same holds true for the so called Kaluza--Klein Dark Matter in theory with universal extra-dimensions. 

Other strategies may provide useful hints, such as the multi-wavelength approach, that consists in the combined analysis of astrophysical spectra at different wavelengths, with the aim of observing e.g. the Synchrotron and Inverse Compton emission produced by secondary electrons produced along with gamma-rays by Dark Matter annihilations  \cite{Profumo:2010ya}. Or the study of the angular power spectrum of gamma-ray anisotropies \cite{Ando:2005xg}, that may allow to identify a Dark Matter contribution to the diffuse gamma-ray background. But even in case of detection, it would probably require a long time before these observations are considered as proof of the existence of Dark Matter, because one would have to exclude an astrophysical origin of the signal. Fortunately, although indirect searches may appear to be not particularly suited to provide incontrovertible evidence for Dark Matter, they have the big advantage of not requiring {\it dedicated} experiments, and that some theoretical models are indeed within the reach of current and upcoming experiments in the next 5--10 years. In absence of these (admittedly optimistic) smoking-gun observations, a convincing case for Dark Matter can be made only in case of successful searches at accelerators or direct detection experiments, in which case indirect searches may still provide useful information on the {\it distribution} of Dark Matter.

\section{Direct detection}

The field of direct detection appears perhaps in a better shape, given the prospects to increase the size, and therefore the sensitivity, of current experiments by at least two orders of magnitude within 5--10 years. The idea is to detect the recoil energy of nuclei struck by Dark Matter particles traveling through a detector, through the measurement of the light, the charge or the phonons produced in the target material by the scattering event. The progress made in this field is rather spectacular: the sensitivity of direct detection experiments  has gone down by more than 3 orders of magnitude in the last 20 years \cite{Gaitskell}. Despite the extraordinary technological progress, however, Dark Matter has not been identified yet.

Something similar to the case of indirect detection actually happens for direct searches: there are intriguing signals that might be interpreted in terms of Dark Matter, but the case for WIMPs is simply not strong enough to convince the community. The most well-known and most discussed example is the DAMA/LIBRA experiment, that reported the detection of  a yearly modulation in the measured event rate, compatible with what expected in common Dark Matter models, where the rate is modulated by the Earth's revolution around the Sun \cite{Bernabei:2008yi}. However, its interpretation in terms of the elastic scattering of a WIMP with a mass around $ 10-100 \mbox{ GeV} $ and $ \crosssection \sim 10^{-3}-10^{-5}\, \mbox{pb} $ has been challenged by other experiments \cite{Fornengo}.
The CoGeNT collaboration also recently reported an excess of low-energy events which could be explained in terms of a very light WIMP \cite{Aalseth:2010vx}, but this interpretation is in tension with the first XENON-100 results \cite{Aprile:2010um} (the debate on the DAMA/LIBRA and CoGeNT results is still open \cite{DDdisc}). Finally, the CDMS-II collaboration recently announced the detection of two events compatible with a WIMP signal, still far from a discovery, since the expected background was of 0.8 events \cite{Ahmed:2009zw}. 

What can we expect from future experiments, and how do we get convincing evidence? The first obvious step would be the detection of a rate of events significantly larger ( i.e. 5 standard deviations) than the expected background, as determined before the unblinding of the data. One can then try to assess the WIMP mass and scattering cross sections compatible with the measured rate. Given the small number of events of the first detection, the reconstruction procedure will likely be rather poor, unless the mass of the Dark Matter particle is very small\cite{Green:2008rd}. 

At that point, in order to validate the Dark Matter interpretation, it will be crucial to add an independent piece of evidence, such as discovery in accelerators, as we shall see in the next section, or direct detection in a different experiment, better if with a different target material, which would actually provide a strong support to the first discovery claim, allow a much more precise determination of the 
WIMP mass \cite{Drees:2008bv}, and an effective way to discriminate among WIMP candidates \cite{Bertone:2007xj}. 

In fig. 1 we summarize the current situation of direct Dark Matter searches. The figure shows the sensitivity of current and upcoming experiments, compared with theoretical predictions, in the wimp-proton cross-section. The cross-section shown in figs. 1 and 2 corresponds to a scalar (or spin-independent) coupling. WIMPs can also interact with the spin of the nucleon, with an axial (or spin-dependent) coupling. The theoretical predictions depend on the specific model considered, and most often they concern different incarnation of the aforementioned neutralino, the Dark Matter candidate that arises from Supersymmetric models. 

The stars, for instance, correspond to a set of benchmark models in a Supersymmetric theoretical setup called Minimal Supergravity (mSUGRA) \cite{Battaglia:2003ab} (see Box 2 for further details on theoretical models). Another very similar Supersymmetric setup is the constrained Minimal Supersymmetric Model; the red contours show the most probable region of the parameter space of this theoretical setup, as determined with a Monte Carlo Markov Chain procedure \cite{Trotta:2008bp}. It is worth noticing that these prediction have been made in the framework of a constrained version of a more general class of supersymmetric models, that allow in general a much more rich phenomenology: the blue contours show the result of a similar Markov Chain scan in the framework of a more general Supersymmetric model with 7 free parameters \cite{Baltz:2004aw} (see Box 2). 

As one can see, a big portion of the parameter space where theoretical models lie will be probed by ton-scale experiments that are expected to start operations within 5--10 years. This is good news, as for this set of parameters we can perform the program described above. One has however to consider the possibility that Supersymmetry, or in general the Dark Matter particle, is outside the reach of ton-scale experiments. In this case the question will arise of whether one should continue searching, and build even bigger and more expensive detectors, or simply stop, and focus on something different. The answer will likely depend on what is found in accelerators, as we shall see in the next section. But it is worth recalling that neutrinos coherent interactions provide an irreducible background for these searches, therefore limiting the capability to probe very low scattering cross sections \cite{strigari}.

\section{Accelerators}
The detection strategy that appears perhaps most promising today is the search for new physics in accelerators. There are in fact big expectations for the Large Hadron Collider, that has recently started operations at CERN. The current plan is to run it at a center of mass energy of 7 TeV until the end of 2011, and then, after an upgrading procedure, at 14 TeV. This will allow to test the existence of new particles at the TeV scale, i.e. where most theorists believe that signs of new physics should appear. 

\paragraph{Box 2: Beyond the Standard Model.} {\it The Standard Model (SM) of particle physics is viewed by many as an effective field theory valid for energies {\it up to} the TeV scale, rather than a truly fundamental theory. This belief is not based on discrepancies with experimental results, but on (very strong) theoretical arguments. Among them, the so-called {\it hierarchy problem} is perhaps the most prominent: in order to stabilize the mass of the Higgs against quadratically divergent radiative corrections without an unacceptable amount of fine-tuning (an adjustment of 32 orders of magnitude, for the SM to be valid up to the Planck scale), the scale of new physics must be ${\cal O} (1)$ TeV, i.e. within the reach of the LHC (see e.g.  \cite{Ellis:2010kf}). As reasonable and aesthetically appealing as it is, this is not  a rigorous mathematical  argument, and the actual scale of new physics could be in principle even higher, depending on the amount of fine-tuning one is willing to tolerate. 

Among the proposed extensions of the Standard Model, Supersymmetry is undoubtedly the most studied, and probably one of the best motivated. The so called Minimal Supersymmetric Standard Model has however about 120 free parameters, and although not all of them are relevant for the calculation of the properties of Dark Matter candidates, some assumptions must be made on the structure of the theory in order to reduce the number of free parameters, and make quantitative predictions for the mass and couplings of supersymmetric particles. In this article we refer to some of the most popular supersymmetric models: 
\begin{itemize}
\item the constrained Minimal Supersymmetric Model and Minimal Supergravity are supersymmetric theories with 4 free parameters. They differ in some small technical details, but both are both often used to make predictions for Dark Matter searches, because despite the very strong theoretical assumptions made to reduce the number of free parameters (i.e. universality of masses and couplings at the Grand Unification scale), they capture the main aspects of the phenomenology of Supersymmetric theories. 
\item the phenomenological Supersymmetric Model is a phenomenological model that is specifically tailored to the study of Dark Matter. There are different versions of the model, one of the most popular being a 7 free parameters theory, where all parameters are specified at low energy, as implemented in the popular DarkSUSY code \cite{darksusy}. A less constrained version of this model has 24 free parameters, and it is the one adopted to discuss the complementarity of direct and accelerator searches (see in particular Figure 2).
\end{itemize}}

Among the proposed extensions of the Standard Model, Supersymmetry is undoubtedly the most studied, and probably one of the best motivated. Not only it solves in a natural way the hierarchy problem, but it also provides a perfect Dark Matter candidate (more than one, in fact \cite{Ellis:2010kf}). There are large portions of the Supersymmetric parameter space within the reach of the LHC, and there are good chances to discover it at the LHC within 5--10 years \cite{Nath:2010zj}.
There is no need to stress the impact that a detection of new physics would have our description of the Universe. We limit ourselves to discuss here the consequences for Dark Matter searches. In particular, a natural question to ask is: how do we understand whether the newly discovered particles have something to do with the Dark Matter in the Universe? From accelerator measurements, in fact, we can infer the existence of a particle which is stable over the timescale it takes for it to escape the detector, i.e. less than 1$\mu$s. But we cannot prove that it is stable over {\it cosmological} timescales, nor we can assess its relic density in absence of a theoretical framework in which to perform the calculation of the cosmological evolution of its density. 

Even if Supersymmetry is discovered, and the mass spectrum of new particles is determined with good accuracy, reconstructing the relic density of the neutralino will be challenging \cite{Baltz:2006fm}, unless one performs the analysis in a low-dimensional parameter space (e.g. mSUGRA \cite{Nath:2010zj}). In this {\it dream scenario}, fortunately, particle astrophysics experiments can provide complementary information on the nature of Dark Matter. In fact, direct searches provide an effective way to reduce degeneracies in the parameter space of new theories, when reasonable assumptions are made on the distribution of Dark Matter particles in the Milky Way. We show in figure \ref{fig:relic_density} an example of a recent study in the framework of a 24-parameters Supersymmetric setup (see Box 2), where the simulated response of the LHC and of 1-ton experiments to a given benchmark model was used to reconstruct the relic density of Dark Matter, showing that a convincing identification of Dark Matter particles is possible with a combination of LHC and direct detection data\cite{Bertone:2010rv}. 

\section{The future}
The other possibility is of course that new physics is {\it not} found at the LHC within 5--10 years. For the reasons we have discussed above, null searches at the LHC would push the scale of new physics into more and more unnatural territory (i.e. to high levels of fine-tuning), and although they would not rule out Supersymmetry and many other new theories, they would cast doubts on the very existence of new physics at any scale, especially if the Higgs boson is found, completing the Standard Model. 

For WIMP Dark Matter studies, the consequences would be dramatic. In absence of new colliders, for which in any case we would have to await at least for a couple of decades,  the only remaining hope would be to obtain smoking-gun evidence from direct or indirect detection. But indirect searches are complicated, as we have seen, and even assuming that one can make a strong case for e.g. Supersymmetry at a higher scale, they are actually much more difficult for high-mass Dark Matter particles, due to the fact that the annihilation spectra scale with the inverse of the mass {\it squared}, and that in general the detection of photons and anti-matter is difficult at energies above tens of TeV. As for direct detection, in absence of any trace of new physics at the LHC, it will be probably difficult to motivate the construction of experiments beyond the ton scale. In absence of any signal we would be left with the {\it nightmare Dark Matter scenario} of null searches at the LHC, direct and indirect detection experiments, a circumstance that would likely mark the decline of the WIMPs, in favor of alternative explanations, such as axions or alternative theories of gravity, provided that they can be reconciled with lensing observations. 

Let us stay optimistic, though. The plans to detect Dark Matter in the near future have been laid out carefully, and they deserve to be carried out with the outmost care, as a discovery would mark the start of a new era of physics, and it would represent the best reward to decades of painstaking searches.


\begin{addendum}
 \item[Competing Interests] The authors declare that they have no
competing financial interests.
 \item[Correspondence] Correspondence and requests for materials
should be addressed to bertone@iap.fr
\end{addendum}


\begin{figure}
\caption{\label{fig:DD} Title: Status of direct Dark Matter searches. Legend: Comparison of theoretical predictions for the strength of WIMP interactions with ordinary matter, with the sensitivity of current and upcoming direct detection experiments. The figure shows the status of direct Dark Matter searches in the scalar WIMP-proton scattering cross-section versus WIMP mass plane. Current experiments have excluded models above the solid lines (CDMS II, green solid line, and Xenon100, black solid). The reach of several upcoming experiments is shown by the dashed lines (from top to bottom, reach of SuperCDMS Phase C with 1 ton of Germanium, LUX with 3 tons of Liquid Xenon, and Xenon1T with 1 ton of Xenon). Also shown for comparison are the predictions for different theoretical models. Stars correspond to benchmarks models (corresponding to typical regions in the theoretical parameter space) in a constrained supersymmetric setup with only 4 free parameters (see text for further details, in particular Box 2) \cite{Battaglia:2003ab}. The most probable region of the parameter space for this theoretical setup can be determined with a Monte Carlo Markov Chain procedure \cite{Trotta:2008bp} , and it is shown here in red. Finally, we also show the result of a scan of the parameter space performed in a less constrained Supersymmetric setup with 7 free parameters specified at low-energy \cite{Baltz:2004aw}, to stress the fact that a more rich phenomenology is in general allowed by Supersymmetric theories. The plot has been made with DMTools \cite{dmtools}.}  \epsfig{file=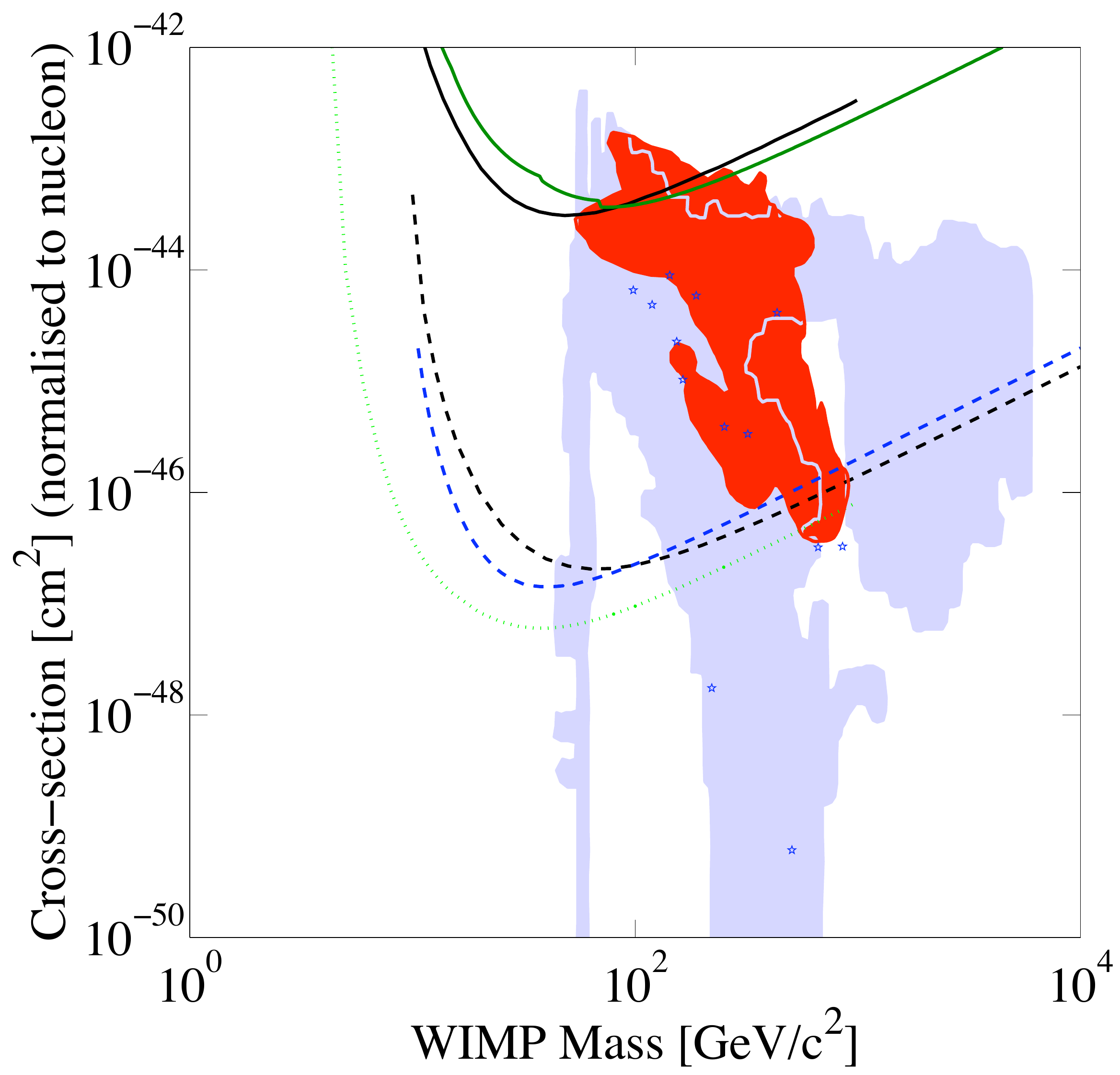,width=0.75\textwidth} 
\end{figure}

\begin{figure}
\caption{\label{fig:relic_density} Title: complementarity between accelerator and direct detection searches. Legend: This figure illustrates the complementarity between accelerator and direct detection searches. We show a reconstruction of the properties of the Dark Matter particle in the scattering cross section vs. relic density plane, starting from the benchmark point indicated by the yellow/red diamond (see text for further details). This model is within the reach of the Large Hadron Collider, so one can simulate the set of measurements that should become available with, say, 300 fb$^{-1}$ of data, corresponding roughly to the data that will be accumulated by 2016, if the experiment runs according to the plans. The result of the reconstruction procedure based on LHC data only, as performed in a supersymmetric setup with 24 free parameters (see Box 2), is shown by the light grey contours, which exhibit a double peak structure, with a very broad peak around the true value. Fortunately, in case of direct detection with a ton-scale experiment, the reconstruction procedure becomes much more precise, as shown by the filled colored contours, since this type of experiment breaks the degeneracy in the parameter space along the dashed line. In this case the best fit point, shown by the encircled black cross, practically lies on top of the true value \cite{Bertone:2010rv}.}  
\epsfig{file=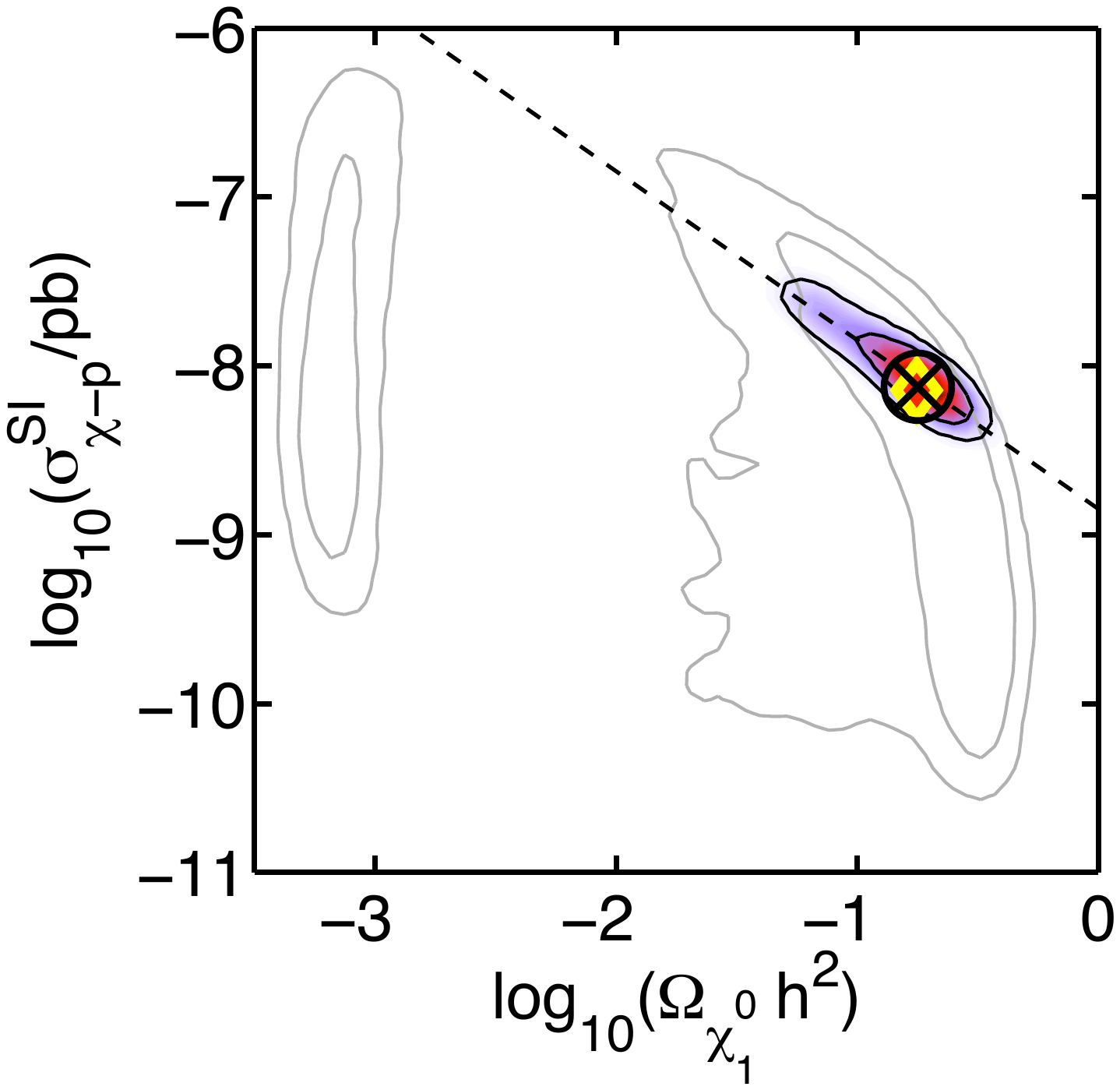,width=0.95\textwidth} 
\end{figure}


\end{document}